\documentclass[a4paper,10pt]{article}

\usepackage[english]{babel}
\usepackage[T1]{fontenc}
\usepackage[dvips]{graphicx}
\usepackage{amsmath, amsfonts,amsthm, mathrsfs, amssymb}
\usepackage{verbatim}

\usepackage{enumerate}
\usepackage[latin1]{inputenc}
\usepackage{geometry}

\def\sidjd{(1+| i^d-j^d|)}

\def\sgeq{\succeq}
\def\vf{\varphi}

\def\eX{e^{-\im\epsilon X}}
\def\meX{e^{\im\epsilon X}}

\def\radE{\sqrt{E-V(q)}}
\def\rady{\sqrt{1-\vy}}

\def\sq{\sqrt{1-|y|^{2l}}}

\def\vy{\tilde V_E(y)}
\def\ir{{\rm i}}

\def\diag{{\rm diag}}

\def\norma#1{\left\|#1\right\|}
\def\sleq{\preceq}

\def\hz{h_0}

\newcommand{\C}{{\mathbb C}}

\newcommand{\R}{{\mathbb R}}

\newcommand{\T}{{\mathbb T}}
\newcommand{\Z}{{\mathbb Z}}

\newcommand{\cA}{{\mathcal A}}
\newcommand{\cB}{{\mathcal B}}

\newcommand{\cH}{{\mathcal H}}

\newcommand{\cO}{{\mathcal O}}

\newcommand{\cS}{{\mathcal S}}
\newcommand{\cT}{{\mathcal T}}

\newcommand{\norm}[1]{\| #1 \|}

\newcommand{\im}{{\rm i}}

\def\uno{{\bf 1}}

\numberwithin{equation}{section}
\newtheorem{theorem}{Theorem}[section]
\newtheorem{lemma}[theorem]{Lemma}
\newtheorem{corollary}[theorem]{Corollary}

\newtheorem{definition}[theorem]{Definition}
\newtheorem{remark}[theorem]{Remark}

\title{Reducibility of 1-d Schr\"odinger equation with time quasiperiodic
  unbounded perturbations, II}

\author{D. Bambusi\footnote{Dipartimento di Matematica, Universit\`a degli Studi di Milano, Via Saldini 50, I-20133
Milano. \newline
 \textit{Email: } \texttt{dario.bambusi@unimi.it}},
}

\begin{document}

\maketitle

\begin{abstract}
We study the Schr\"odinger equation on $\R$ with a potential behaving
as $x^{2l}$ at infinity, $l\in[1,+\infty)$ and with a small time
  quasiperiodic perturbation. We prove that, if the perturbation
  belongs to a class of unbounded symbols including smooth potentials
  and magnetic type terms with controlled growth at infinity, then the
  system is reducible. 
\end{abstract}

\section{Introduction}

The present paper is a continuation of \cite{1} in which a
reducibility result for the  time dependent
Schr\"odinger equation
\begin{align}
\label{schro}
\ir \dot\psi&=(H_0+\epsilon W(\omega t))\psi\ , \ x\in\R
\\
\label{H}
&H_0=-\partial_{xx}+V(x)\ ,
\end{align}
with $W$ a suitable unbounded perturbation was proved. The improvement
we get here is that we deal with a more general class of
perturbations. For example we prove here reducibility, if $V(x)\simeq
|x|^{2l}$, $l\geq1$, as $x\to\infty$, and
\begin{equation}
\label{W}
W(\omega t)=a_0(x,\omega t)-\ir a_1(x,\omega t)\partial_x\ ,
\end{equation}
with $a_i$ functions of class $C^\infty$ fulfilling
\begin{align}
\label{a0}
\left|\partial_x^ka_0(x,\omega t)\right|&\sleq \langle
x\rangle^{\beta_2-k}\ ,\quad \beta_2<l\ ,
\\
\label{a1}
\left|\partial_x^ka_1(x,\omega t)\right|&\sleq \langle
x\rangle^{\beta_3-k}\ ,\quad \left\{
\begin{matrix}
\beta_3<l-1&{\rm if}&1<l\leq 2
\\
\beta_3<{l}/{2}&{\rm if}&2<l
\end{matrix}
\right.\ ;
\end{align}
in the case $l=1$, $a_1$ must vanish identically. The theory developed
in \cite{1} only allowed to deal with the case of polynomial $a_0$ and
$a_1$, but a faster growth at infinity of both $a_0$ and $a_1$ was
allowed. 

As usual, boundedness of Sobolev norms and pure
point nature of the Floquet spectrum follow. 

We recall that previous results on the reducibility problem for
perturbations of the Schr\"odinger equation have been obtained in
quite a number of papers for the superquadratic case with bounded or
unbounded perturbations (see in particular
\cite{DS96,DSV02,BG01,LY10,EK09}); in the quadratic case the only
available results deal with bounded perturbations
\cite{C87,W08,GT11,GP16}. The result of the present paper allows a
growth of the perturbation at infinity faster then all the previous
papers dealing with the one dimensional case (except \cite{1}). On the
other hand, we assume here that $W$ is a symbol with the property that
its derivatives of suffitiently high order decay fast at infinity
(essentially as in \eqref{a0},\eqref{a1}); this is not required in
most papers on reducibility. Concerning the higher dimensional case,
it is not clear if the present method can be extended in order to
deal with it.

The idea of the proof (following \cite{PT01,BBM14}, see also
\cite{Mtesi,FP15,BM16a}) is to use pseudo-differential calculus in order
to conjugate the original system to a system with a smoothing
perturbation and then to apply KAM theory. In the present paper we
just prove the smoothing result, since afterwards one can
apply the KAM type theorem of \cite{1} in order to conclude the
proof. From the technical point of view the result is obtained by
introducing a new class of symbols. However, when working with such a
class it becomes quite complicated to show that the function used to
generate the smoothing transformation is actually a symbol. The proof
of this property occupy the majority of the paper. We also would like
to mention that the class of symbols we use is a variant of the class
introduced by Hellfert and Robert in \cite{HR82}.

\noindent{\it Acknowledgments}. This paper originated from a series
of discussions with quite a lot of people. In particular I warmly
thank P. Baldi, R. Montalto and M. Procesi who explained to me in a
quite detailed way their works. During the preparation of the present
work I benefit of many suggestions and discussions with A. Maspero and
D. Robert. In particular D. Robert pointed to my attention (and often
explained me) the papers \cite{HR82,HR82D}. I also thank
B. Gr\'ebert for some relevant discussions on the Harmonic case.

\section{Statement of the Main Result}\label{main}

Fix a real number $l\geq 1$ and define the weights
\begin{equation}
\label{lam}
\lambda(x,\xi):=\left(1+\xi^2+|x|^{2l}\right)^{1/2l}\ ,\quad \langle x\rangle:=\sqrt{1+x^2}
\end{equation}
\begin{definition}
\label{Sm}
The space $S^{m_1,m_2}$ is the space of the symbols $g\in C^\infty(\R)$ such
that $\forall k_1,k_2\geq 0$ there exists $C_{k_1,k_2}$ with the
property that
\begin{equation}
\label{sm}
\left|\partial^{k_1}_\xi\partial^{k_2}_{x}g(x,\xi)\right|\leq
C_{k_1,k_2} \left[\lambda(x,\xi)\right]^{m_1-k_1l}\langle x\rangle^{m_2-k_2}\ .
\end{equation}
The best constants $C_{k_1,k_2}$ such that \eqref{sm} hold form a
family of seminorms for the space $S^{m_1,m_2}$. 
\end{definition}

To a symbol $g\in S^{m_1,m_2}$ we associate its Weyl quantization, namely
the operator $g^w(x,D_x)$, $D_x:=-\ir \partial_x$, defined by
\begin{equation}
\label{weyl}
G\psi(x)\equiv g^w(x,D_x)\psi(x):=\frac{1}{2\pi}\int_{\R^2}e^{\ir(x-y)\cdot
  \xi}g\left(\frac{x+y}{2};\xi\right) \psi(y)dyd\xi\ .
\end{equation} 
{\it We will denote by a capital letter the Weyl
  quantized of a symbol denoted with the corresponding lower case
  letter.} The only exception will be the perturbation $W$ (we mainly
think of it as a potential).

In the following we will denote by
$\cS^{m_1,m_2}:=C^{\infty}(\T^n,S^{m_1,m_2})$ the space of
$C^{\infty}$ functions on $\T^n$ with values in $S^{m_1,m_2}$.
The frequencies $\omega$ will be assumed to vary in the set 
$$
\Omega:=[1,2]^{n}\ ,
$$
or in suitable closed subsets
$\widetilde\Omega$. 

We denote by $S^{m_1,m_2}_N$ the space of the symbols which are only $N$ times
differentiable and fulfill the inequality \eqref{sm} only for
$k_1+k_2\leq N$. This is a Banach space with the norm
\begin{equation}
\label{Sn}
\norm{g}_{S^{m_1,m_2}_N}:=\sum_{k_1+k_2\leq
  N}\sup_{(x,\xi)\in\R^2}\frac{\left|\partial^{k_2}_x
  \partial_{\xi}^{k_1}g(x,\xi)
  \right|}{[\lambda(x,\xi)]^{m_1-lk_1}\langle x\rangle^{m_2-k_2}}\ .
\end{equation}

We remark that for the space $\cS^{m_1,m_2}$ a family of seminorms is given by
the standard norms of $C^M(\T^n;S^{m_1,m_2}_N)$ as $M$ and $N$ vary.

In the case $l>1$, the potential $V$ defining $H_0$ is assumed to belong to
$S^{0,2l}$ to be symmetric, namely
\begin{equation}
\label{symV}
V(x)=V(-x)\ ,
\end{equation}
and furthermore to admit an asymptotic expansion of the
form
\begin{equation}
\label{quasi.1}
V(x)\sim|x|^{2l}+\sum_{j\geq 1} V_{2l-2j}(x)
\end{equation}
with $V_k$ homogeneous of degree $k$, namely s.t., $V_k(\rho
x)=\rho^kV(x)$, $\forall \rho>0$. 

\noindent
We also assume that 
\begin{align}
\label{V3}
V'(x)\not=0\ ,\quad \forall x\not=0\ .
\end{align}

\begin{remark}
\label{r.v.1}
The assumptions \eqref{symV}, \eqref{quasi.1} are used in order to
simplify the proofs of Lemmas \ref{lememd} and \ref{lemchi}; they can
probably be relaxed. Assumption \eqref{V3} can also be weakened in
order to deal with the case where the set of the critical points of
$V$ is bounded.
\end{remark}

\noindent
An example of a non-polynomial potential fulfilling the assumptions is
$$
V(x)=\langle x\rangle^{2l}\ .
$$

In the case $l=1$ we assume that 
$$
V(x)=x^2\ .
$$

The unperturbed Hamiltonian $H_0$ is the quantization of the
classical Hamiltonian system with Hamiltonian function
\begin{equation}
\label{h0c}
h_0(x,\xi):=\xi^2+V(x)\ .
\end{equation}
\begin{remark}
\label{periodic}
As a consequence of the assumptions above all the solutions of the
Hamiltonian system $h_0$ are periodic with a period $T(E)$ which
depends only on $E=h_0(x,\xi)$. 
\end{remark}

We will denote by $\Phi^t_{h_0}$ the flow of the Hamiltonian system
\eqref{h0c}.

We denote by $\lambda_j^v$ the sequence of the eigenvalues of $H_0$.
In what follows we will identify $L^2$ with $\ell^2$ by introducing
the basis of the eigenvector of $H_0$. 

We use the symbol $\cA(x,\xi):=(1+h_0(x,\xi))^{\frac{l+1}{2l}}$ to
{\it define, for $s\geq 0$, the spaces $\cH^s=D([\cA^w(x,-\ir
    \partial_x)]^{s})$ (domain of the $s$- power of the operator
  operator $\cA^w(x,-\ir \partial_x)$) endowed by
  the graph norm}. For negative $s$, the space $\cH^s$ is the dual of
$\cH^{-s}$.

We will denote by $B(\cH^{s_1};\cH^{s_2})$ the space of bounded linear
operators from $\cH^{s_1}$ to $\cH^{s_2}$.

In order to state the assumptions on the perturbation we define the
average with respect to the flow of $h_0$:
\begin{equation}
\label{mediaW}
\langle
W\rangle(x,\xi,\omega t):=\frac{1}{T(E)}\int_0^{T(E)}W(\Phi^\tau_{h_0}(x,\xi),\omega
t)d\tau\ ;
\end{equation}
then, for $m\in\R$, we  denote 
\begin{equation}
\label{quad}
[m]:=\max\left\{0,m\right\}\ .
\end{equation}

Concerning the perturbation, we assume that $W\in
\cS^{\beta_1,\beta_2}$ and we define
\begin{equation}
\label{betatilde}
\tilde\beta:=\left\{
\begin{matrix}
2\beta_1+[\beta_2]+[\beta_2-1]-2l+1 & {\rm if} \quad\langle W\rangle\equiv
0 \ {\rm and}\ l>1
\\
\beta_1+[\beta_2] & {\rm otherwise}
\end{matrix}
\right.\ .
\end{equation}

\begin{theorem}
\label{m.1}
Assume $$\tilde\beta<l\quad {\rm and }\quad
\beta_1+[\beta_2]<2l-1\ ,$$ then there exists $\epsilon_*>0$ and
$\forall \left|\epsilon\right|<\epsilon_*$ a closed set
$\Omega(\epsilon)\subset\Omega$ and, $\forall
\omega\in\Omega(\epsilon)$ there exists a unitary (in $L^2$) time
quasiperiodic map $U_\omega(\omega t)$ s.t. defining  $\varphi$  by
$U_\omega(\omega t)\varphi=\psi $, it satisfies the equation
\begin{equation}
\label{rido}
\ir \dot \varphi= H_{\infty}\varphi\ ,
\end{equation}  
with $H_{\infty}=\diag (\lambda_j^\infty)$, with
$\lambda_j^\infty=\lambda_j^\infty(\omega,\epsilon)$ independent of time and 
\begin{equation}
\label{dla}
\left|\lambda_j^\infty-\lambda_j^v\right|\leq C\epsilon
j^{\frac{\tilde \beta}{\tilde l+1}}\ ,
\end{equation}
for some positive $C$. Furthermore one has
\begin{itemize}
\item[1.]
  $\displaystyle{\lim_{\epsilon\to0}\left|\Omega-\Omega(\epsilon)\right|=0}$; 
\item[2.] $\forall s,r\geq0$, $\exists \epsilon_{s,r}>0$ and $s_r$
  s.t., if $|\epsilon|<\epsilon_{s,r}$ then the map $\phi\mapsto
  U_\omega(\phi)$ is of class $C^r(\T^n;B(\cH^{s+s_r};\cH^{s}))$; in
  particular one has $s_0=0$ and $s_1=\beta_1+[\beta_2]$.
\item[3.]  $\exists b>0$ s.t. $\forall |\epsilon|<\epsilon_{s,1}$, one
  has
  $\norma{U_\omega(\phi)-\uno}_{B(\cH^{s+\beta_1+[\beta_2]};\cH^{s})
  }\leq C_s\epsilon^b$.
\end{itemize}
\end{theorem}

\begin{remark}
\label{r.m.2}
If $W$ is the sum of different addenda, then 
Theorem \ref{m.1} applies also if its assumptions are fulfilled by each
of the addenda separately. This is particularly relevant in the case
where the average of some of the addenda vanishes. Thus in this case
the value of $\tilde \beta$ can depend on the addendum one is
considering.    
\end{remark}

\begin{corollary}
\label{c.m.1}
If $W$ is given by \eqref{W}, then Theorem \ref{m.1} applies under the
conditions \eqref{a0} and \eqref{a1}. 
\end{corollary}
\proof The condition on $\beta_2$ is obvious. Consider the addendum
$-\ir a_1(x,\omega t)\partial_x$, which has symbol 
$$
a_1(x,\omega t)\xi+S^{0,\beta_3-1}\ ,
$$ and remark that, by Eq. \eqref{media} below, the average of the
main term vanishes and therefore for this term $\tilde \beta$ is given
by the first of \eqref{betatilde} which is made explicit by \eqref{a1}.  \qed

\begin{remark}
\label{r.m.1}
In the case of the quartic oscillator ($l=2$) and perturbation of the
form \eqref{W}, we have the bounds $\beta_2<2$ and $\beta_3<1$. We
recall that \cite{LY10} had $\beta_2\leq 1$ and $\beta_3\leq0$, but
the perturbation was not assumed to be asymbol. In
\cite{1} we were able to deal also with some cases with $\beta_2=4$
and $\beta_3=2$, but only when $a_0,$ and $ a_1$ are polynomial.

We also remark that the assumption that the functions $a_i$ are
symbols rules out cases like $a_i(x,\omega t)=\cos(x-\omega t)$.
\end{remark}
\begin{remark}
\label{harm}
In the case of the Harmonic oscillator we cover the perturbations of
the class considered in \cite{W08} (in which the decay at infinity of
$a_0$ and its derivatives are exponential) and in the counterexample
of \cite{D14}.

On the contrary the perturbations in \cite{GT11} (which must decay at
infinity) and in \cite{GY00}
can belong to a class of symbols in which the decay at infinity does
not improve as one extracts derivatives.
\end{remark}


\section{Proof of Theorem \ref{m.1}}\label{egor}

\subsection{Some symbolic calculus}\label{symbol}

First we remark that $S^{m_1,m_2}\subset S^{m_1+[m_2],0}$.

In the proof we will also need the classes of symbols used in
\cite{1}, thus we recall the corresponding definitions

\begin{definition}
\label{Sm.1}
The space $S^{m}$ is the space of the symbols $g\in C^\infty(\R)$ such
that $\forall k_1,k_2\geq 0$ there exists $C_{k_1,k_2}$ with the
property that
\begin{equation}
\label{sm.1}
\left|\partial^{k_1}_\xi\partial^{k_2}_{x}g(x,\xi)\right|\leq
C_{k_1,k_2} \left[\lambda(x,\xi)\right]^{m-k_1l-k_2}\ .
\end{equation}
\end{definition}

In order to deal with functions $p$ such that
there exist a $\tilde p$ with the property that
$$
p(x,\xi)=\tilde p(h_0(x,\xi))\ ,
$$
we introduce the following class of symbols. 

\begin{definition}
\label{d.sm}
A function $\tilde p\in\C^{\infty}$ will be said to be of class
$\widetilde S^m$
if one has
\begin{equation}
\label{sm.12}
\left|\frac{\partial^k\tilde p}{\partial E^k}(E)\right|\sleq \langle
E^{\frac{m}{2l}-k}\rangle \ .
\end{equation} 
\end{definition}

By abuse of notation, we will
say that $p\in\widetilde S^m$ if there exists $\tilde p\in \widetilde
S^m$ s.t. $p(x,\xi)=\tilde p(h_0(x,\xi))$.

We will also need to use functions from $\T^n$ to $\widetilde
S^m$. The corresponding class will be denoted by $\widetilde \cS^{m}$.

We now give a reformulation of the results of sect. 4.1 of \cite{1} in
the case of the symbols of the classes $S^{m_1,m_2}$.

The application of the Calderon Vaillencourt theorem yields the
following Lemma.

\begin{lemma}
\label{caderon}
Let $f\in S^{m_1,m_2}$, then one has
\begin{equation}
\label{CV}
f^w(x,D_x)\in B(\cH^{s_1+s};\cH^{s})\ ,\quad \forall s\ ,\quad \forall
s_1\geq m_1+[m_2]\ .
\end{equation}
\end{lemma}

Given a symbol $g\in S^{m_1,m_2}$ we will write
\begin{equation}
\label{asym}
g\sim\sum_{ j\geq 0}g_j\ ,\quad g_j\in
S^{m_1^{(j)},m_2^{(j)}}\ ,\quad m_1^{(j)}+[m_2^{(j)}  ]\leq
m_1^{(j-1)}+[m_2^{(j-1)} ]\ ,
\end{equation}
if $\forall \kappa$ there exist $N$ and $r_N\in
S^{-\kappa,0}$, s.t.
$$
g=\sum_{j=0}^{N}g_j+r_N\ .
$$

\begin{lemma}
\label{l.M}
Given a couple of symbols $a\in S^{m_1,m_2}$ and $b\in S^{m_1',m_2'}$, denote by
$a^w(x,D_x)$ and $b^w(x,D_x)$ the corresponding Weyl
operators, then there exists a symbol $c$, denoted by $c=a\sharp b$ such
that
$$
(a\sharp b)^{w}(x,D_x)=a^w(x,D_x)b^w(x,D_x)\ ,
$$ 
furthermore one has
\begin{equation}
\label{sharp}
(a\sharp b)\sim \sum_{j\geq 0} c_j
\end{equation}
with 
$$
c_j=\sum_{k_1+k_2=j}\frac{1}{k_1!k_2!}\left(\frac{1}{2}\right)^{k_1}
\left(-\frac{1}{2}\right)^{k_2}  (\partial^{k_1}_\xi D^{k_2}_xa)
(\partial^{k_2}_\xi D^{k_1}_xb)\in S^{m_1+m_1'-lj,m_2+m_2'-j} \ .
$$
In particular we have
\begin{equation}
\label{moy}
\left\{a;b\right\}^q:=-\ir(a\sharp b-b\sharp
a)=\left\{a;b\right\}+S^{m_1+m_1'-3l,m_2+m_2'-3}\ ,
\end{equation}
where 
$$
\left\{a;b\right\}:=-\partial_\xi a\partial_xb+\partial_\xi
b\partial_xa\in S^{m_1+m_1'-l,m_2+m_2'-1}\ ,
$$
is the Poisson Bracket between $a$ and $b$, while \eqref{moy} means
that $\left\{a;b\right\}^q=\left\{a;b\right\}+$some quantity belonging
to  $ S^{m_1+m_1'-3l,m_2+m_2'-3}$. 
\end{lemma}

\begin{definition}
\label{pseud}
An operator $F$ will be said to be a pseudo-differential operator of
class $O^{m_1,m_2}$ if there exists a sequence $f_j\in
S^{m_1^{(j)},m_2^{(j)}}$ with $m_1^{(j)}+[m_2^{(j)}]\leq
m_1^{(j-1)}+[m_2^{(j-1)}] $ and, for any $\kappa$ there exist $N$ and
an operator $R_N\in B(\cH^{s-\kappa};\cH^{s})$, $\forall s$ such that
\begin{equation}
\label{expa}
F=\sum_{j\geq0}^Nf_j^w+R_N\ .
\end{equation}

 In this case we will write $f\sim\sum_{j\geq 0} f_j$ and $f$ will be said to
 be the symbol of $F$; the function $f_0$ will be said to be the
 principal symbol of $F$.
\end{definition}

Concerning maps we will use the following definition

\begin{definition}
\label{pseudomap}
A map $\T^n\ni\phi\mapsto F(\phi)\in O^{m_1,m_2}$, will be said to be
smooth of class $\cO^{m_1,m_2}$ if the functions of the sequence $f_j$
also depend smoothly on $\phi$, namely
$f_j\in\cS^{m_1^{(j)},m_2^{(j)}}$ and the operator valued map
$\phi\mapsto R_N(\phi)$ has the property that for any $K\geq1 $ there
exists $a_K\geq 0$ s.t. for any $N$ one has
\begin{equation}
\label{propcN}
R_N(.)\in C^K(\T^n;B(\cH^{s-\kappa+a_K};\cH^{s}))\ , \forall s \ .
\end{equation} 
\end{definition}

Finally we need (Whitney) smooth functions of the
frequencies. Following \cite{1} (and \cite{Stein}), we will denote by
$Lip_\rho(\widetilde\Omega;\cB)$ the functions of
$\omega\in\tilde\Omega$ with values in a Banach space $\cB$ which have
$k$ derivatives of H\"older class $\rho-k$. Here $k$ is the first
integer strictly smaller then $\rho$ and $\widetilde \Omega\subset
\Omega$ is a closed set.

\begin{definition}
\label{why.1}
We will say that a function $f:\widetilde\Omega\to\cS^{m_1,m_2}$ is of
class $Lip_\rho^{m_1,m_2}(\widetilde\Omega)$ if forall $N_1,N_2$ it is of
class
$Lip_\rho(\widetilde\Omega;C^{N_1}(\T^n;S^{m_1,m_2}_{N_2}))$. Similarly
we will say that $f\in \widetilde{Lip}_\rho^{m}(\widetilde\Omega)$ if
forall $N_1,N_2$, one has 
$f\in Lip_\rho(\widetilde\Omega;C^{N_1}(\T^n;\widetilde S^{m}_{N_2}))$.
\end{definition}

\subsection{Quantum Lie transform}\label{qliet}

Given a symbol $\chi$, we consider the corresponding Weyl operator
$X$. If $X$ is selfadjoint, then we will consider the unitary operator
$e^{-\ir \epsilon X}$. The following Lemma gives a sufficient
condition for selfadjointness.

\begin{lemma}
\label{MR}
Let $\chi\in S^{m,0}$ have the further property that
$\partial_x\chi\in S^{m-1,0}$. Assume $m\leq l+1$, then
$X:=\chi^w(x,D_x)$ is selfadjoint and $e^{-\ir \epsilon X}$ leaves
invariant all the spaces $\cH^s$.
\end{lemma}
\proof We use Proposition A.2 of \cite{MR16}. To ensure the result it
is enough to exhibit a positive selfadjoint operator $K$ such that
both the operators $XK^{-1}$ and $[X,K]K^{-1}$ are bounded. To this
end we take $K$ to be the Weyl operator of the symbol
$\cA:=(1+h_0)^{\frac{l+1}{2l}}\in S^{l+1}$. From symbolic calculus it
follows that $XK^{-1}\in O^{0,0}$ which is thus bounded and, by the
additional property on the $x$ derivative of $\chi$, one has
$\{\chi;\cA \}\in S^{2m-l-1,0}$ so that $[X,K]K^{-1}\in O^{m-l-1,0}$,
which is bounded under the assumption of the Lemma.\qed

Next we use the operator  $e^{-\ir \epsilon X}$ to transform
operators. 

\begin{definition}
\label{lie}
Let $X$ be a selfadjoint operator; we will say that 
\begin{equation}
\label{qlie}
(Lie_{\epsilon X}F):=\meX F\eX  
\end{equation}
is the quantum Lie transform of $F$ generated by $\epsilon X$.
\end{definition}
It is easy to see that defining 
\begin{equation}
\label{serqlie}
F_0=F\ ;\quad F_k:=-\ir [F_{k-1};X]\ ,
\end{equation}
one has
\begin{equation}
\label{r.lie}
\frac{d^k}{d\epsilon^k}Lie_{\epsilon X}F=\meX F_k\eX\ .
\end{equation}
and therefore (formally)
\begin{align}
\label{qlie1}
 Lie_{\epsilon X}F =\sum_{k\geq 0}\frac{1}{k!}\epsilon^k F_k\ .
\end{align}
We will use these formulae in situations where the series are
asymptotic. 

We will use the same terminology also when $X$ depends on time and/or
on $\omega$ (which in this case play the role of parameters).

We are interested in the way Hamiltonian operators change their
form in the case where $X$ also depends on time. The following Lemma
is Lemma 3.2 of \cite{1} to which we refer for the proof.

\begin{lemma}
\label{T.1}
Let $F$ be selfadjoint operator which can also depend on time, and let
$X(t)$ be a family of selfadjoint operators smoothly dependent on
time. Assume that $\psi(t)$ fulfills the equation
\begin{equation}
\label{1}
\im\dot \psi=F\psi\ ,
\end{equation}  
then $\vf$ defined by 
\begin{equation}
\label{2}
\vf=e^{\im\epsilon X(t)}\psi\ ,
\end{equation}
fulfills the equation 
\begin{equation}
\label{3}
\im\dot \vf =F_\epsilon(t) \vf 
\end{equation}
with 
\begin{align}
\label{4.1.1}
F_\epsilon&:=Lie_{\epsilon X} F-Y_X\ ,
\\
\label{yx}
&Y_X:=\int_0^\epsilon (Lie_{(\epsilon-\epsilon_1)X}\dot X)d\epsilon_1\ .
\end{align}
\end{lemma}

In the case where both $F$ and $X$ are pseudo-differential operators
one can reformulate everything in terms of symbols.  Thus, if $f$ and
$\chi$ are symbols and $\chi$ fulfills the assumptions of Lemma
\ref{MR} one can define
\begin{equation}
\label{liqs}
f_0^q:=f\ ,\quad f_k^q:=\left\{ f_{k-1}^q;\chi\right\}^q\ ,
\end{equation}
and one can expect the symbol of $Lie_{\epsilon X}F$ to be
$\sum_{k\geq 0}\epsilon^kf_k^q/k!$. A sufficient condition is given by
the following lemma:

\begin{lemma}
\label{egorov}
Let $\chi\in S^{m,0}$ and let
$f\in S^{m_1,m_2}$ be symbols, assume $m<l$,
then $Lie_{\epsilon X}F\in O^{m_1,m_2}$, and furthermore its symbol, denoted by
$lie_{\epsilon\chi} f$, fulfills
\begin{equation}
\label{liqser}
lie_{\epsilon \chi} f\sim \sum_{k\geq
  0}\frac{\epsilon^kf_k^q}{k!} \ .
\end{equation}
\end{lemma}
\proof First remark that $f^q_k\in
S^{m_1+k(m-l),m_2-k}$. From \eqref{r.lie} and the formula
of the remainder of the Taylor expansion one has
$$
Lie_{\epsilon X}F=\sum_{k=0}^{N}\frac{F_k}{k!}\epsilon^k+\frac{\epsilon^{N+1}}{N!}
\int_0^1(1+u)^J e^{-\ir u\epsilon X}F_{N+1}e^{\ir
  u\epsilon X}du\ ,
$$
so that, by defining $R_N$ to be the integral term of the previous
formula, we have $R_N\in B(\cH^{s-\kappa},\cH^s)$ with
$\kappa=N(l-m)-m-[-N+m_2]$, which diverges as $N\to\infty$
and thus shows that the expansion \eqref{liqser} is asymptotic in the
sense of definition \ref{pseud}.\qed 

\begin{remark}
\label{yx.1}
Let $\chi\in S^{m,0}$ be such that $\partial_x \chi\in S^{m-1,0}$,
with $m<l$, then
the operator $Y_X$ defined by
eq. \eqref{yx} is a pseudo-differential operator of class $O^{m,0}$ with
symbol 
\begin{equation}
\label{yx.2}
y_x:=\int_0^\epsilon(lie_{(\epsilon-\epsilon_1)\chi}\dot\chi)d\epsilon_1=
\dot\chi +\epsilon S^{2m-l-1,0}\ . 
\end{equation}
\end{remark}

\subsection{Main lemmas}\label{l.s.1}
The algorithm used in order to conjugate the original system to a
system with a smoothing perturbation is the one described in Sect. 4.2
of \cite{1}. In order to make it effective in the present case we have
to prove that the solutions of the homological equations
are symbols. In this sub section we present the homological equations
and give the Lemmas solving them; they will be used in the proof of
the smoothing theorem which will be given in the next subsection. The
proof of these lemmas is the main technical result of the paper and
will be given in Sect. \ref{coho}.

From now on we will use the notation 
\begin{equation}
\label{sleq}
a\sleq b
\end{equation}
to mean ``there exists a constant $C$ independent of all the relevant
quantities, such that $a\leq C b$''.

\vskip 5pt
As the example of the period $T(E)$ in the case $V(x)=x^{2l}$ (with
$l$ integer) shows, it is useful to deal with functions which have a
singularity at zero. In order to avoid the problems it creates we will
first regularize the functions at zero and solve the homological
equations only outside a neighborhood of zero.

\vskip10pt

The first homological equation we have to solve is the following one
\begin{equation}
\label{12}
p+\left\{h_0;\chi\right\}=\langle p\rangle\ ,
\end{equation}
where $\langle p\rangle$ is defined by \eqref{mediaW} with $p$ in
place of $W$. The problem is to determine $\chi$ s.t. \eqref{12}
holds. 

First we have the following Lemma.

\begin{lemma}
\label{lememd}
Let $p\in S^{m_1,m_2}$ be a symbol supported outside a neighborhood
of zero (in the phase space), then $\langle p\rangle$ is a symbol of
class $ \widetilde S^{m_1+[m_2]}$ and is supported outside a
neighborhood of zero.
\end{lemma}

Concerning the solution of the homological equation we have the
following Lemma.

\begin{lemma}
\label{lemchi}
Let $p\in S^{m_1,m_2}$ be a symbol supported outside a neighborhood
of zero, then the homological equation
\eqref{12} has a solution $\chi$ which is a symbol of class $\chi\in
S^{m_1+[m_2]-l+1,0}$ with the further property that $\partial_x
\chi\in S^{m_1+[m_2]-l,0}$ and is supported outside a neighborhood
of zero.
\end{lemma}

\begin{remark}
\label{funzioni}
In the above lemmas $p$ can also depend on the angles $\phi$ and on
the frequencies $\omega$, but they only play the role of parameters,
so in that case the result is still valid substituting the classes
$\cS$ or $Lip_\rho$ with the same indexes to the classes $S$.
\end{remark}

In order to iterate the procedure, when $l>1$, we will have to solve an
equation of the form of \eqref{12} with $h_0$ replaced by 
\begin{equation}
\label{h1}
h_1:=h_0+\epsilon f(h_0)\ ,
\end{equation}
with $f\in \tilde S^{m}$ and $m<l$, namely equation
\begin{equation}
\label{12.3}
p+\left\{h_1;\chi\right\}=\langle p\rangle\ ,
\end{equation}

\begin{lemma}
\label{lemchi.3}
Let $l>1$ and $p\in S^{m_1,m_2}$ be a symbol supported outside a neighborhood
of zero, then the homological equation \eqref{12.3} has a solution
$\chi$ which is a symbol of class $\chi\in S^{m_1+[m_2]-l+1,0}$ and
$\partial_x \chi\in S^{m_1+[m_2]-l,0}$.
\end{lemma}

The third homological equation we have to solve is 
\begin{equation}
\label{12.p}
-\omega\cdot\frac{\partial \chi}{\partial\phi}=p-\bar p\ ,
\end{equation}
where $p$ is a symbol and $\bar p$ is defined by
\begin{equation}
\label{media.2}
\bar p(x,\xi):=\frac{1}{(2\pi)^{n}}\int_{\T^n}p(x,\xi,\phi)d\phi\ .
\end{equation}
Such an equation was already studied in \cite{1} and the solution was
obtained in Lemma 4.20 of that paper which is already in the form we
need in the present paper. We now give its statement (for the proof we
refer to \cite{1}).

Fix $\tau>n-1$ and denote 
\begin{equation}
\label{diof}
\Omega_{0\gamma}:=\left\{\omega\in\Omega\ :\ \left|k\cdot\omega\right|\geq\gamma
|k|^{-\tau}\right\}\ ,
\end{equation}
then it is well known that 
\begin{equation}
\label{sti}
\left|\Omega-\Omega_{0\gamma}\right|\sleq \gamma\ .
\end{equation}

\begin{lemma}
\label{coh2}
Let $p\in \widetilde{Lip}_\rho^{m}(\Omega_{0\gamma})$, be a symbol,
then there exists a solution $\chi\in
\widetilde{Lip}_\rho^{m}(\Omega_{0\gamma}) $ of Eq. \eqref{12.p}. Furthermore $\bar p\in
\widetilde{Lip}_\rho^{m}(\Omega_{0\gamma}) $.
\end{lemma}

Finally, in the case of the Harmonic oscillator $l=1$, we will meet the
following homological equation 
\begin{equation}
\label{12.1}
\left\{h_0,\chi\right\}-\dot\chi+p=\overline{\langle p\rangle}\ .
\end{equation}

In order to solve it, define the set 
\begin{equation}
\label{tigamma}
{\Omega_{1\gamma}}:=\left\{\omega\in\Omega\ :\ \left|\omega\cdot
k+k_0|\geq\frac{\gamma}{1+|k|^\tau}\right|\ ,\ (k_0,k)\in\Z^{n+1}-\left\{0\right\}
\right\}\ .
\end{equation}

\begin{lemma}
\label{lemchi1}
Let $p\in Lip_\rho^{m_1,m_2}(\Omega_{1\gamma})$, then there exists a
solution $\chi\in Lip_\rho^{m_1+[m_2],0}(\Omega_{1\gamma}) $ of
\eqref{12.1}. Furthermore $\overline{\langle p\rangle}\in
\widetilde{Lip}_\rho^{m_1+[m_2]}(\Omega_{1\gamma})$. 
\end{lemma}

\subsection{The smoothing theorem and end of the proof of Theorem
  \ref{m.1}}\label{smot} 

\begin{theorem}
\label{smoothing}
Fix $\gamma>0$ small, $\rho>2$ and an arbitrary $\kappa>0$. Assume
\begin{align}
\label{sti.1}
\beta_1+[\beta_2]< 2l-1\quad {\rm and}\quad \tilde \beta<l
\end{align} 
then there exists a (finite) sequence of symbols $\chi_1,...,\chi_N$
with $\chi_j\in Lip_\rho^{m^{(j)}_1,m^{(j)}_2}(\Omega_{0\gamma})$,
$m^{(j)}_1+[m^{(j)}_2] \leq \beta_1+[\beta_2]$ $\forall j$, s.t., defining 
\begin{equation}
\label{Xj}
X_j:=\chi^w_j(x,D_x,\omega t)\ ,\quad \omega\in\Omega_{0\gamma}\ ,
\end{equation}
such operators are selfadjoint and the transformation
\begin{equation}
\label{transfe}
\psi=e^{-\im\epsilon  X_1(\omega t)}....e^{-\im\epsilon X_N(\omega t)}\varphi\ ,
\end{equation}
transforms $H_\epsilon(\omega t)$ (c.f. \eqref{H}) into a
pseudo-differential operator $H^{(reg)}$ with symbol $h^{(reg)}$ given
by
\begin{equation}
\label{hreg}
h^{(reg)}=h_0+\epsilon z+\epsilon \widetilde{z}+\epsilon r
\end{equation}
where $z\in \tilde S^{\tilde \beta}$ is a function of $h_0$ independent of
time and of $\omega$; $\tilde z\in \widetilde{Lip}_\rho^{2\tilde\beta-2l+1
}(\Omega_{0\gamma})$ is an $\omega$ dependent function of $h_0$
independent of time, and $r$ depends on
$(x,\xi,\phi,\omega)$. Furthermore one has
\begin{align}
\label{reg.3.1}
r&\in Lip_\rho^{-\kappa,0}(\Omega_{0\gamma})\ .
\end{align}
In the case $l=1$ the set $\Omega_{0\gamma}$ must be substituted by
the set $\Omega_{1\gamma}$.
\end{theorem}

\noindent 
{\it Proof of Theorem \ref{smoothing} in the case $l>1$}.
Denote
$$
\beta:=\beta_1+[\beta_2]\ ,\quad m:=\beta-l+1\ .
$$ 
Let $\eta$ be a $C^\infty$ function such that
\begin{equation}
\label{cutoff}
\eta(E)=\left\{   
\begin{matrix}
1 & {\rm if} & |E|>2
\\
0 & {\rm if} & |E|<1
\end{matrix}
\right.
\end{equation}
and split 
\begin{equation}
\label{Wsplit}
W=W_0+W_\infty\ ,\quad W_\infty(x,\xi)=W(x,\xi)(1-\eta(h_0(x,\xi)))\ ,\quad
W_0(x,\xi)=W(x,\xi)\eta(h_0(x,\xi))\ , 
\end{equation}
then $W_\infty\in \cS^{-\kappa_1,-\kappa_2}$ for any
$\kappa_1,\kappa_2$, and $W_0\in \cS^{\beta_1,\beta_2}$ is the actual perturbation
that has to be transformed into a regularizing operator.

The proof of the smoothing theorem is based only on the solution of
the homological equation and the computation of symbols of commutators,
which (up to operators which are smoothing of all orders) are
operations preserving the property of symbols of being zero in the
region $E<1$.

So, we forget $W_{\infty}$ and transform $h_0+\epsilon W_0$ using the
operator $X_1$ with symbol $\chi_1$ obtained by solving the
homological equation \eqref{12} with $p=W_0$, so that $\chi_1\in
\cS^{m,0}$, with $\partial_x\chi_1\in \cS^{m-1,0}$ so that by Lemma
\ref{MR} the corresponding Weyl operator is selfadjoint provided $m
\leq l+1$ and Lemma \ref{egorov} applies provided $m<l$ (implied by
\eqref{sti.1}).

Then the symbol of the
transformed Hamiltonian is given by
\begin{align}
\label{h1.a}
h^{(1)}&:=h_0+\epsilon(\langle
W_0\rangle-W_0)
+\epsilon \cS^{m-l,-3}+\epsilon^2
S^{\beta+m-l-1,0}+\epsilon^2S^{\beta_1+m-l,\beta_2-1} 
\\
&+\epsilon W_0+\epsilon^2  S^{\beta_1+m-l,\beta_2-1}
\\
&-\epsilon \dot \chi_1+\epsilon^2 S^{2m-(l+1),0}
\\
\label{h1.b}
&=h_0+\epsilon\langle W_0\rangle-\epsilon\dot\chi_1 +\epsilon p_1\ ,
\end{align}
with $p_1\in S^{\beta+m-l-1,0}+S^{\beta_1+m-l,\beta_2-1}$.

Consider first the case where $\langle W_0\rangle\equiv 0$. In this case
 we determine $\chi_2$ by solving the homological equation
\eqref{12} with $p_1$ in place of $p$, A simple analysis shows that 
$$
\langle p_1\rangle\in \tilde S^{2\beta_1+[\beta_2]+[\beta_2-1]-2l+1}\equiv
\tilde S^{\tilde\beta} \ ,\quad \chi_2\in S^{\tilde\beta-l+1,0}\ .
$$
Since $\tilde \beta <l$, $lie_{\epsilon\chi_2}$ has the property that,
if $f\in S^{m_1,m_2}$, then
\begin{equation}
\label{err}
lie_{\epsilon\chi_2}f-f\in \sum_{j} S^{m_1^{(j)},m_2^{(j)}}\ ,\quad
m_1^{(j)}<m_1\ {\rm and}\quad m_2^{(j)}<m_2\ .
\end{equation} 
Thus, the transformed Hamiltonian has the form
\begin{equation}
\label{tildeh1}
 h^{(1_2)}=h_0+\epsilon\langle p_1\rangle-\epsilon \dot
\chi_1+l.o.t
\end{equation}
where l.o.t. means terms with the property analogue to
\eqref{err}. Next we eliminate $-\dot \chi_1$. To this end we
determine $\chi_3$ by solving \eqref{12} with $-\dot \chi_1$ in place
of $p_1$. Remark that $\langle\dot \chi_1\rangle\equiv 0$ so that
$\chi_3\in S^{\beta_1+[\beta_2]-2l+2,0}$ transforms $h^{(1_2)}$ into
$$
h^{(1_3)}:=h_0+\epsilon\langle p_1\rangle-\epsilon\dot \chi_3+l.o.t\ .
$$
Then (if needed) we iterate again until we get 
$$ \tilde h^{(1)}=h_0+\epsilon\langle p_1\rangle+\epsilon
\sum_jS^{\beta_1^{(j)},\beta_2^{(j)}}\ ,
$$
with $\beta_1^{(j)}+[\beta_2^{(j)}]< \tilde \beta$, $\forall j$.

Thus, both in the case $\langle W_0\rangle=0$ and in the case  $\langle
W_0\rangle\not=0$, we are reduced to a
Hamiltonian of the form 
\begin{equation}
\label{htra1}
h^{(1')}:=h_0+\epsilon f(h_0,\omega t)+\epsilon p_2\ ,
\end{equation}
with $f(h_0,.)\in \cS^{\tilde \beta}$ and $p_2$ a lower order
correction in the above sense.

We now continue, following \cite{1}, by eliminating the time dependence
from $f$. Thus take $\chi_4$ to be the solution of Eq. \eqref{12.p}
with $p=f(h_0)$, so that $\chi_4\in \widetilde{Lip}_\rho^{\tilde
  \beta}(\Omega_{0\gamma})$. Provided $$\tilde \beta<l\ ,$$ one gets
that the corresponding Weyl operator is selfadjoint and the quantum
lie transform it generates, transforms symbols into symbols and has
the property \eqref{err}. Then the
symbol of the transformed Hamiltonian takes the form
\begin{align*}
h^{(2)}=& h_0+\epsilon\overline{f(h_0)}+\epsilon p_2+ l.o.t.
\end{align*}
where all the functions are defined on $\Omega_{0\gamma}$ and 
$$
p_2\in \sum_{j} S^{\beta_1^{(j)},\beta_2^{(j)}}\ ,\quad
\beta_1^{(j)}+[\beta_2^{(j)}] <\tilde \beta-l\ . 
$$
In particular the l.o.t. is the
lowest order term with a nontrivial dependence on $\omega$.

Denote now
$$h_1:=h_0+\epsilon f(h_0)
$$ 
and iterate the construction with $h_1$ in place of $h_0$. At each
step of the iteration one gains $l$, in the sense that one passes from
a perturbation (of a time independent Hamiltonian) which belongs to
some classes $S^{\tilde \beta_1,\tilde\beta_2}$ to perturbations
belonging to classes  $S^{\tilde \beta_1',\tilde\beta_2'}$ with
$$
\tilde \beta_1'+[\tilde \beta_2']\leq \tilde \beta_1+[\tilde
  \beta_2]-l \ . 
$$
Thus the result follows.
\qed

\noindent
{\it Proof of Theorem \ref{smoothing} in the case $l=1$.} First remark
that  $\beta<1$ implies $\beta_1<1$ and $\beta_2<1$. 
We make a first step by taking
$\chi_1\in  Lip_\rho^{\beta}$
to be the solution of Eq. \eqref{12.1} with $p=W$. Remarking that in this case,
for any symbol $f$, one has 
$$
\left\{h_0,f\right\}^q=\left\{h_0,f\right\}\ ,
$$
it follows that the transformed Hamiltonian is 
$$
h^{(1)}=h_0+\epsilon\overline{\langle W\rangle}+\epsilon^2 r_1\ ,
$$
with 
$$
r_1\in  Lip_\rho^{2\beta-2,0}+ Lip_\rho^{\beta+\beta_1-1,0}\subset L
ip_\rho^{\beta^{(1)},0}\ ,\quad \beta^{(1)}:=\beta+\beta_1-1\ .
$$
Then we iterate getting 
$$
h^{(2)}=h_0+\epsilon\overline {\langle W\rangle}+\epsilon^2 \overline
{\langle r_1\rangle} +\epsilon^3r_2\ ,
$$ with $r_2\in
Lip^{\beta+\beta^{(1)}-2,0}+Lip^{\beta^{(1)}+\beta^{(1)}-1,0}$. If
$\beta-2>\beta^{(1)}-1$ the dominant term is the first one and we put
$\beta^{(2)}:= \beta^{(1)}-2+\beta$, otherwise we define
$\beta^{(2)}:= 2\beta^{(1)}-1$. Thus in particular we have
$\beta^{(2)}<\beta^{(1)}$. Then we iterate and at each step we get a
remainder $r_N\in Lip^{\beta^{(N)},0}$, with a sequence $\beta^{(N)}$
diverging at $-\infty$. We remark that, after some steps, one will get 
$\beta-2>\beta^{(N)}-1$, and therefore, from such a step one will have
simply $\beta^{(N+1)}=\beta^{(N)}-2+\beta$.

Finally we remark that the average of $r_1$ is the first term in the
time independent part which depends on $\omega$.
\qed

After the  smoothing
Theorem \ref{smoothing}, the Hamiltonian of the system is reduced to
the form \eqref{hreg} to which we apply the methods (and the results)
of \cite{1}. Precisely using, Lemmas 5.1 and 5.2 and Corollary 5.4 of
\cite{1} one has the following Lemma
\begin{lemma}
\label{Hop}
For any $\gamma>0$ and $\rho\geq 2$ there exists a positive
$\epsilon_*$ s.t., if $|\epsilon|< \epsilon_*$ then there exists a set
$\Omega^{(0)}_\gamma$, and a unitary (in $L^2$) operator $U_1$
Whithney smooth in $\omega\in\Omega^{(0)}_\gamma$, fulfilling
\begin{align}
\label{diaga.301}
\left|\Omega-\Omega^{(0)}_\gamma\right|\sleq \gamma^a
\\
\label{U2}
U_1^*H^{(reg)} U_1=A^{(0)}+\epsilon R_0\ ,
\end{align}
where $a$ is a positive constant (independent of $\gamma,\epsilon$). 
The operator $A^{(0)}$ is given by
\begin{align}
\label{A}
A^{(0)}:=\diag(\lambda_j^{(0)})\ ,
\end{align}
with $\lambda_j^{(0)}=\lambda_j^{(0)}(\omega)$ Whitney smooth in
$\omega$ fulfilling the following inequalities
\begin{align}
\label{diaga.1}
\left|\lambda_j^{(0)}-\lambda_j^v\right|\sleq j^{\frac{\tilde
    \beta}{l+1}}\ ,
\\
\label{diaga.2}
\left|\lambda_i^{(0)}-\lambda_j^{(0)}\right|\sgeq
\left|i^d-j^d\right|\ ,
\\
\label{diaga.3}
\left|\frac{\Delta(\lambda_i^{(0)}-\lambda_j^{(0)})}{\Delta\omega}
\right| 
\sleq
\epsilon|i^d-j^d| \ .
\\
\label{diaga.303}
\left|\lambda_i^{(0)}-\lambda_j^{(0)}+\omega\cdot k\right|\geq
\frac{\gamma\sidjd}{1+|k|^\tau}\ ,\quad
\left|i-j\right|+\left|k\right|\not=0\ ,
\end{align}
where, as usual, for any Lipschitz function $f$ of $\omega$, we
denoted $\Delta f=f(\omega)-f(\omega')$.

Furthermore, $\forall s$ $\exists \epsilon_s$, s.t., if
$|\epsilon|<\epsilon_s$ then
\begin{align}
\label{U1}
\norma{U_1-\uno}_{
  Lip_{\rho}(\Omega^{(0)}_\gamma;B(\cH^{s-\delta};\cH^{s})) }\sleq
\epsilon\ ,\quad \delta:=\tilde \beta-(l+1)\ ,
\\
\label{HoR.1}
R_0:=U_1^{-1}RU_1\in Lip_\rho(\Omega^{(0)}_\gamma;
C^\ell(\T^n;B(\cH^{s-\kappa};\cH^s)))\ , \quad \forall \ell\ .
 \end{align}
\end{lemma}

\noindent {\it End of the proof of Theorem \ref{m.1}.} Now Theorem
\ref{m.1} is obtained immediately by applying Theorem 7.3 of \cite{1}
to the system \eqref{U2}.\qed

\section{Proof of the main lemmas}\label{coho} 

In this section we prove Lemmas \ref{lememd}, \ref{lemchi},
\ref{lemchi.3} and \ref{lemchi1}.

To prove that $\langle p\rangle$ and $\chi$ are symbols we
use some explicit formulae for the solution of second order equations
in order to write in a quite explicit form the integrals over the
orbits of $h_0$.

 Consider the Hamilton equations of $\hz$, namely
\begin{equation}
\label{hameq}
\dot \xi=-\frac{\partial V}{\partial x}\ ,\quad \dot x=\xi\ .
\end{equation}
It is well known that one can exploit the conservation of energy in
order to reduce the system to quadrature, namely to compute the time
as a function of the position:
\begin{equation}
\label{time}
t(x,x_0)=\int_{x_0}^x \frac{dq}{\sqrt{E-V(q)}}\ .
\end{equation}
One also has that the period $T(E)$ is given by
\begin{equation}
\label{per}
T(E)=4\int_{0}^{q_M(E)}\frac{dq}{\sqrt{E-V(q)}}\ ,
\end{equation}
where $q_M=q_M(E)$ is the positive solution of the equation 
\begin{equation}
\label{qM.1}
E=V(q_M)\ .
\end{equation}

Before giving the proof of the main Lemmas, we need some preliminary
results. First, in order to compute and estimate integrals of the form
\eqref{time}, \eqref{per}, we will often use the change of variables
\begin{equation}
\label{change.y}
q(y)=q_My\ .
\end{equation}
Furthermore it is useful to define the function 
\begin{equation}
\label{tildev}
\tilde v(E,y):=\sqrt{\frac{1-|y|^{2l}}{1-\frac{V(q(y))}{E}}}\ ,
\end{equation}
so that one has
\begin{equation}
\label{tildev.1}
\frac{1}{\sqrt{1-\frac{V(q(y))}{E}}}=\frac{\tilde
  v(E,y)}{\sqrt{1-|y|^{2l}}}\ .
\end{equation}

\begin{lemma}
\label{qM} The quantity $q_M$
has the form 
\begin{equation}
\label{qM.2}
q_M(E)\sim E^{1/2l}\bar q(E)\ ,
\end{equation}
where the function $\bar q$ admits an asymptotic expansion in powers
of $\mu^2:=E^{-1/l}$ and its first term is $1$.
\end{lemma}
\proof Consider equation \eqref{qM.1}, divide by $E=\mu^{-2l}$; using
the asymptotic expansion \eqref{quasi.1} it
takes the form 
$$ 1\sim \sum_{j\geq 0} \mu^{2l} V_{2l-2j}(q_M)=\sum_{j\geq 0}
\mu^{2l-2j}\mu^{2j} V_{2l-2j}(q_M)= \sum_{j\geq 0}
\mu^{2j} V_{2l-2j}(\mu q_M)=\bar q^{2l}+  \sum_{j\geq 1}
\mu^{2j} V_{2l-2j}(\bar q)\ .
$$ 
Thus one sees that $\bar q$ admits an
asymptotic expansion in powers of $\mu^2$. \qed

\begin{lemma}
\label{tildev.3}
For all $E_0>0$, the function $\tilde v(E,y)$ is a
$C^\infty([E_0,\infty))$ function of $E$ and one has
\begin{equation}
\label{tildev.4}
\left|\frac{\partial^k \tilde v}{\partial E^k}(E,y)\right|\sleq
\frac{1}{E^k}\ ,\quad \forall y\in[-1,1]\ ,\quad \forall E\geq E_0\ .
\end{equation}
\end{lemma}
\proof Denote $\tilde V_E(y):=\frac{V(q(y))}{E}$ and remark that, due
to the definition of $q(y)$, one has $\tilde V_E(\pm1)\equiv 1$, so
that $\tilde v$ is regular at $y=\pm1$. Furthermore, by Lemma
\ref{qM} (and its proof), one has
\begin{equation}
\label{tildev.6}
\tilde V_E(y)\sim \bar q^{2l}|y|^{2l}+\sum_{j\geq 1} \mu^{2j}V_{2l-2j}
(\bar q y)\ ,
\end{equation}
(with $\mu=E^{-1/2l}$) which shows that $\tilde V_E(y)$ admits an
asymptotic expansion in $\mu$. First we remark that, by
eq. \eqref{tildev.6} and Lemma \ref{epM}, the thesis of the Lemma
holds true for $y$ outside a neighborhood of $\pm1$. We discuss now
the result for $y$ near 1.

We use the Faa di Bruno formula in order to compute the derivatives of
$$
\tilde v\equiv \frac{\sqrt{1-|y|^{2l}}}{\sqrt{1-\tilde V_E(y)}}
$$ 
with respect to $E$. Denote $f(x):=(1-x)^{-1/2}$. Remark
that
$$
f^{(j)}(x)=C_j\frac{(1-x)^{-j}}{\sqrt{1-x}}\ ,
$$
and compute 
\begin{align}
\nonumber
\frac{\partial^k }{\partial E^k} f(\tilde V_E)\asymp \sum_{j=1}^{k}
f^{(j)}(\tilde V_E )\sum_{h_1+...+h_j=k} \partial_E^{h_1}\tilde
V_E...\partial_E^{h_j}\tilde V_E
\\
\label{tildev.8}
\asymp
\frac{1}{\sqrt{1-x}}\sum_{j=1}^{k} \sum_{h_1+...+h_j=k}\frac{\partial_E^{h_1}\tilde
V_E}{1-\tilde V_E} ... \frac{\partial_E^{h_j}\tilde
V_E}{1-\tilde V_E} \ .
\end{align} 
We study the single fraction at r.h.s.. Compute the Taylor expansion
of $\tilde V_E(y)$ at $y=1$, it is given by
\begin{equation}
\label{taytilde}
\tilde V_E(y)\simeq 1+\sum_{k\geq1}\frac{1}{E} V^{(k)}(E^{1/2l}\bar
q)(E^{1/2l}\bar q)^k\frac{(y-1)^k}{k!}\ ,
\end{equation}
from which we get 
\begin{equation*}
\frac{\partial_E^{h}\tilde
V_E}{1-\tilde V_E}\simeq
\frac{\sum_{k\geq1}\partial^h_E\left[\frac{1}{E}V^{(k)}(E^{1/2l}\bar 
q)(E^{1/2l}\bar q)^k\frac{(y-1)^{k-1}}{k!}  \right]}{\sum_{k\geq1}
  \frac{1}{E}V^{(k)}(E^{1/2l}\bar 
q)(E^{1/2l}\bar q)^k\frac{(y-1)^{k-1}}{k!} }\ ,
\end{equation*}
which is regular at $y=1$. To get a more usable expression and an
estimate of this fraction we remark that the single term of the sum in
the numerator is a multiple of 
$$
\partial_E^h[\partial_y^k\tilde V_E]_{y=1}=[\partial_y^k\partial_E^h\tilde
  V_E]_{y=1}\ ,
$$ and one can compute the r.h.s. exploiting the asymptotic expansion
\eqref{tildev.6} of $\tilde V_E$. So one gets that $\partial_y\tilde
V_E$ admits an asymptotic expansion in $\mu^2$. Thus one can apply
Lemma \ref{epM} which shows that the single term in the sum in the
numerator of the fraction is estimated by $E^{-(h+1/l)}$. Inserting in
\eqref{tildev.8} one gets the thesis. \qed

\begin{lemma}
\label{period}
The period $T=T(E)$ is s.t. $T\eta\in S^{1-l}$, where $\eta$ is
the cutoff function defined in \eqref{cutoff}.
\end{lemma}
\proof Due to the presence of the cutoff function it is enough to
study the behavior of $T(E)$ at infinity.
Making the change of variables \eqref{change.y} in the integral
\eqref{per}, we get
\begin{equation}
\label{tq}
T=\frac{4q_M}{E^{1/2}}\int_{0}^{1}\frac{dy}{\sqrt{1-\vy}}=\frac{4\bar
  q}{E^{\frac{1}{2}-\frac{1}{2l}}} \int_{0}^{1}\frac{\tilde
  v(E,y)}{\sqrt{1-y^{2l}}} \ ;
\end{equation}
exploiting the property \eqref{tildev.4} of the function $\tilde v$
one immediately gets the thesis.
 \qed
 
We are now ready for proving that the average of a symbol is a symbol.

\noindent {\it Proof of Lemma \ref{lememd}} Remark that $\langle
p\rangle$ is a function of $E$ only. To compute it we first make a
change of variables in the phase space, namely we will use the
variables $(E,x)$ instead of $(x,\xi)$. Such a change of variables is
well defined in the region $\xi>0$ (or $\xi<0$) and for
$-q_M<x<q_M$. In these variables the flow $\Phi_{\hz}$ is given by
$E(t)=E$ and $x(t)$ given by the inverse of the formula
\eqref{time}. Thus, using the definition of the average and making the
change of variables $t(q)$ in the integrals, we have
\begin{equation}
\label{media}
\langle p\rangle (E)
=\frac{1}{T(E)}\int_{-q_M}^{q_E}\frac{p(q,\radE)}{\radE}dq +\frac{1}{T(E)}
\int_{-q_M}^{q_E}\frac{p(q,-\radE)}{\radE}dq  \ .
\end{equation}
Consider the first term (the second one can be treated in the same
way); making the change of variables \eqref{change.y} it takes the form
\begin{equation}
\label{m3.1}
 \frac{q_M}{ T(E) E^{1/2}}\int_{-1}^1\frac{p\left(q(y),E^{1/2}\rady\right)\tilde
   v(E,y)}{\sq}dy \ .
\end{equation}
This quantity and its derivatives with respect to $E$ can be easily
estimate using Lemma \ref{stint.6} and Lemma \ref{stixi}. \qed

We recall a first representation formula
for $\chi$. The next lemma is Lemma 5.3 of \cite{BG93} to which
we refer for the proof (see also Lemma 4.21 of \cite{1}).

\begin{lemma}
\label{solhomo}
The solution of the homological equation \eqref{12} is given by
\begin{align}
\label{coh.2}
\chi=\frac{1}{T(E)}\int_0^{T(E)}t(p-\langle p\rangle)\circ \Phi^t_{\hz} dt\ .
\end{align}
\end{lemma}

To estimate the function $\chi$ we need some more preliminary work.  

\begin{lemma}
\label{lemchi.1}
Let $p$ be a function, denote $\check p:=p-\langle
p\rangle$ and
\begin{align}
\label{ts}
t_S(x):=\int_{-q_M}^x\frac{dq}{\radE}\ ,\quad
t^-_S(x):=\int_x^{q_M}\frac{dq}{\radE}\equiv t_S(-x)\ , 
\\
\label{dmu}
d\mu^+(q):=\frac{\check p(q,\radE)}{\radE}dq\ ,\quad
d\mu^-(q):=\frac{\check p(q,-\radE)}{\radE}dq\ 
\end{align}
($t_S$ is the time taken to go from $-q_M$ to $x$) then, in the
coordinates $(E,x)$ for the upper half plane, the function $\chi$
defined by \eqref{coh.2} is given by
\begin{align}
\label{chif}
\chi(E,x)&=\frac{1}{T(E)}
\int_{-q_M}^{q_M}(t _S(q)d\mu^+(q)+t^-_S(q)d\mu^-(q))  
+\frac{1}{2}\int_{-q_M}^{q_M}d\mu^-(q)
\\
\label{chif2}
&+\int_{-q_M}^{x}d\mu^+(q)     \ .
\end{align}
\end{lemma}
\proof We use again the formula \eqref{time}. In all the integrals $E$
will play the role of a parameter, so we do not write it in the
argument of the functions. We split the interval of integration in
\eqref{coh.2} into three subintervals. For this purpose we
define $t_M(x):=\frac{T}{2}-t _S(x)$, and remark that this is the
time at which a solution starting at $(x,\xi)$ reaches $(q_M,0)$. We
write
$$
[0,T]=[0,t_M(x)]\cup [t_M,t_M+\frac{T}{2}]\cup [t_M+\frac{T}{2},T]\ ,
$$ and we study separately the integrals over the intervals. 

The first integral is given by
\begin{align}
\label{1.4.0}
\int_0^{t_M}t\check p(\Phi^t_{\hz}(x,\xi))dt=
\int_x^{q_M}\frac{t(q,x)\check p(q,\radE)}{\radE} dq
\\
\label{1.4.11}
=\int_x^{q_M}t _S(q)d\mu^+(q) -t _S(x)\int_x^{q_M}d\mu^+(q)\ ,
\end{align}
where of course $t(q,x)$ is defined by \eqref{time}. The 
integral over the second interval is given by
\begin{align}
\label{1.4.1}
&-\int_{q_M}^{-q_M}(\frac{T}{2}-t_S (x)+t_S^-(q))d\mu^-(q)=
\\
\label{1.4.2}
&=\frac{T}{2}\int_{-q_M}^{q_M}d\mu^-(q)-t _S(x)\int_{-q_M}^{q_M}
d\mu^-(q)+\int_{-q_M}^{q_M}t_S^-(q)
d\mu^-(q)\ .
\end{align}
Finally the third integral is given by
\begin{align}
\label{1.4.3}
\int_{-q_M}^x\left[\frac{T}{2}+\left(\frac{T}{2}-t_S (x)\right)+t_S (q)
  \right]d\mu^+ (q) =
\\
\label{1.4.4}
=T\int_{-q_M}^x d\mu^+(q)-t_S (x) \int_{-q_M}^x d\mu^+(q) + 
\int_{-q_M}^xt _S(q) 
 d\mu^+(q) \ .
\end{align}
Summing up we get
\begin{align}
\label{suchi}
\int_{-q_M}^{q_M}(t _S(q)d\mu^+(q)+t^-_S(q)d\mu^-(q))
\\
\label{suchi2}
-t _S(x)\int_{-q_M}^{q_M}(d\mu^+(q)+d\mu^-(q))
\\
+\frac{T}{2}\int_{-q_M}^{q_M}d\mu^-(q)+T\int_{-q_M}^{x}d\mu^+(q)\ ,
\end{align}
but the integral in \eqref{suchi2} is exactly the integral of $\check
p$ along an orbit of $\hz$ and thus it vanishes, thus we get
\eqref{chif} and \eqref{chif2}.  \qed

\begin{lemma}
\label{stichi}
Let $g\in S^{m_1,m_2}$ be a symbol, consider the function 
\begin{equation}
\label{G.def}
G(E,x):=\int_{-q_M}^x\frac{g(q,\sqrt{E-V(q)})}{\sqrt{E-V(q)}}dq\ ,
\end{equation}
and the function  
$$
\widehat G(x,\xi):=G(\xi^2+V(x),x)\ .
$$ Then $\eta(h_0)\widehat G\in S^{m_1+[m_2]-l+1,0}$ and
$\eta(h_0)\partial_x\widehat G\in S^{m_1+[m_2]-l,0}$.
\end{lemma}
\proof Due to the presence of the cutoff function, it is enough to
study the behavior of $\widehat G$ as $E\to\infty$. First we estimate
the modulus of $G$ (and of $\widehat G$). To this end it is better to
represent the integral in terms of integral over the flow of
$h_0$. Preliminarly remark that
\begin{equation}
\label{stichi.100}
\left|g(x,\xi)\right|\sleq \lambda^{m_1}(x,\xi)\langle
x\rangle^{m_2}\sleq \lambda^{m_1+[m_2]}(x,\xi)\sleq \langle
h_0(x,\xi)\rangle^{m_1+[m_2]} \ .
\end{equation}
Using the notation \eqref{time} one has
\begin{align*}
\left|G(E,x)\right|=\left|\int_0^{t_S(x)}g(\Phi^t_{h_0}(-q_M,0))dt\right|\sleq 
\int_0^{T/2}\langle h_0(\Phi^t_{h_0}(-q_M,0))\rangle^{m_1+[m_2]}dt
\\
=
\frac{T}{2} \langle E\rangle^{\frac{m_1+[m_2]}{2l}}\sleq
\lambda^{m_1+[m_2]-l+1}\ .
\end{align*}

To compute the derivatives of $G$ and of $\widehat G$ it is better to
use the formula \eqref{G.def}, to make the change of variables
\eqref{change.y} and to use the function $\tilde v$ defined in
\eqref{tildev}, so that one gets
\begin{equation}
\label{stichi.101}
G(E,x)=\frac{\bar q}{E^{\frac{1}{2}-\frac{1}{2l}}}\int_{-1}^{\frac{\mu
    x}{\bar q}} \frac{\tilde v(E,y)g(q(y),\sqrt{E-V(q(y))})}{\sq} dy
\end{equation}
with $\mu=E^{-1/2l}$. From this formula one can easily compute 
\begin{align}
\label{stichi.102}
\partial_EG&=\partial_E\left( \frac{\bar
  q}{E^{\frac{1}{2}-\frac{1}{2l}}} \right) \int_{-1}^{\frac{\mu
    x}{\bar q}} \frac{\tilde v(E,y)g(q(y),\sqrt{E-V(q(y))})}{\sq} dy
\\
\label{stichi.103}
&+ E^{\frac{1}{2l}}\bar
  q
 \frac{g(x,\sqrt{E-V(x)})}{\sqrt{E-V(x)}}\partial_E\left(\frac{\mu
   x}{\bar q}\right) 
\\
\label{stichi.104}
&+  \frac{\bar
  q}{E^{\frac{1}{2}-\frac{1}{2l}}} \int_{-1}^{\frac{\mu
    x}{\bar q}} \frac{\partial_E\tilde v(E,y)\ g(q(y),\sqrt{E-V(q(y))})}{\sq} dy
\\
\label{stichi.105}
&+  \frac{\bar
  q}{E^{\frac{1}{2}-\frac{1}{2l}}} \int_{-1}^{\frac{\mu
    x}{\bar q}} \frac{\tilde v(E,y)\partial_E
  g(q(y),\sqrt{E-V(q(y))})}{\sq} dy\ ,
\end{align}
where, in order to simplify \eqref{stichi.103} we used the definition
of $\tilde v$. 

Remark now that one has
\begin{equation}
\label{widehat}
\frac{\partial \widehat G}{\partial x}=\frac{\partial G}{\partial
  E}V'+\frac{\partial G}{\partial x}\ .
\end{equation}
We study the contribution of \eqref{stichi.103} to
$\partial\widehat G/\partial x$, which is the most singular one. To
this end we compute
\begin{align}
\label{sti.4}
\frac{\partial G}{\partial x}+V'(x)\,
\eqref{stichi.103}=\frac{g(x,\xi)}{\xi} 
\left[1+q_M V'(x) \partial_E\left(
\frac{ x}{q_M}\right) \right]\ ,
\end{align}
where, when explicitly possible we introduced the variables $(x,\xi)$. 
We study now the square bracket in \eqref{sti.4} in order to show
that \eqref{sti.4} is regular on the line $\xi=0$; we denote by
\begin{equation}
\label{sti.12}
\cT(E,x):= q_M V'(x) \partial_E\left(
\frac{x}{q_M}\right)  \ 
\end{equation}
the second term in the bracket and we simplify it. First remark that
the line $(x,\xi)=(x,0)$, in terms of the variables $(E,x)$, becomes
the curve $(V(x),x)$, which can also be parametrized by $E$ and in such
a parametrization has the form $(E,q_M(E))$. Expanding at
$\xi=0$, one has
\begin{equation}
\label{sti.14}
\widehat
\cT(x,\xi):=\cT(\xi^2+V(x),x)=\cT(V(x),x)+\partial_E\cT(V(x),x)2\xi+O(\xi^2)=
\cT(E,q_M)+2\partial_E\cT(E,q_M)\xi+O(\xi^2)\ .
\end{equation}
Now,
using \eqref{sti.12} and the definition of $q_M$, 
one gets
$$
\cT(E,q_M)=-V'(q_M)\partial_E(q_M)=-V'(q_M)\frac{1}{V'(q_M)}=-1\ .
$$ Inserting in \eqref{sti.14} and substituting in \eqref{sti.4} one
sees that \eqref{sti.4} is regular at $\xi=0$.

In conclusion we have 
\begin{align}
\label{stichi.108}
\partial_x\widehat G&=V'(x)
\frac{E^{\frac{1}{2}}}{q_M}\partial_E\left(
\frac{q_M}{E^{\frac{1}{2}}} \right) \widehat G(x,\xi)
\\
\label{stichi.109}
&+g(x,\xi)\left[\frac{1+\widehat\cT(x,\xi)}{\xi}\right]
\\
\label{stichi.110}
&+V'(x) 
  q_M \int_{-q_M}^{x} \frac{(\partial_E\tilde
    v)(E,y(q))\ g(q,\sqrt{E-V(q)})}{\tilde v(E,y(q))\sqrt{E-V(q)}} dq
\\
\label{stichi.111}
&+ V'(x) 
  q_M \int_{-q_M}^{x} \frac{\partial_E
  g(q(y),\sqrt{E-V(q(y))})}{\sqrt{E-V(q(y))}} dy\ .
\end{align}
Remark that \eqref{stichi.108} and \eqref{stichi.111} clearly have the
same structure as $\widehat G$, so these terms are suitable to start
an iteration which shows that the original quantity is a symbol. One
has still to deal with the other two terms. We start by
\eqref{stichi.109}. 

The analysis of the square bracket in \eqref{stichi.109} (the only
nontrivial part) has to be done by analyzing separately a
neighborhood of $\xi=0$. Such a region can be analyzed by exploiting
the expansion \eqref{sti.14}, which allows to show that it is a symbol
in such a neighborhood. The other region is trivial since the
function is smooth in that region. Doing the explicit computations one
easily shows that it is a symbol. 

We come to \eqref{stichi.110}. We wrote it in that form, since
exploiting it one can compute its derivative with respect to $x$. An
explicit computation shows that, mutatis mutandis, such a derivative
is given again by \eqref{stichi.108}-\eqref{stichi.111}. The main
difference is that \eqref{stichi.109} has to be substituted by 
$$
\frac{g(x,\xi)\partial_E\tilde v(E,x/q_M)}{\tilde
  v(E,x/q_M)}\left[\frac{1+\widehat\cT(x,\xi)}{\xi}  \right]\ ,
$$
which is again a symbol. 

To conclude the proof we estimate the different terms of
\eqref{stichi.108}-\eqref{stichi.111}. The estimate of all the
terms, but \eqref{stichi.109} is obtained by the same argument used to
estimate $G$ which gives that all such terms are bounded by $\langle
x\rangle^{2l-1}\lambda^{m_1+[m_2]-3l+1}$. 

In order to estimate \eqref{stichi.109} we consider its main term in
the expansion in inverse powers of $E$:
$$
\cT(E,x)=V'(x)\left[-x\frac{\partial_E
    E^{1/2l}}{E^{1/2l}}\right]= -\frac{V'(x)x}{2l E}\simeq
-\frac{|x|^{2l}}{E}\ ,
$$ 
so that 
$$
\left|\frac{1+\cT(E,x)}{\xi}\right|\simeq \left|\frac{E-|x|^{2l} }{\xi
E} \right|= \left|\frac{\xi}{E}\right|\sleq \lambda^{-l}\ .
$$
It follows that 
$$
\left|\eqref{stichi.109}\right|\sleq \lambda^{m-(l-1)-1}\ .
$$
\qed

\noindent{\it Proof of Lemma \ref{lemchi}}. First remark that, from
Lemma \ref{stichi}, $\eta t_{S}\in S^{-l+1,0}$ and $\eta \partial_x
t_{S}\in S^{-l,0}$. It follows that $\eqref{chif}\eta\in \tilde S^{m_1+[m_2]-l+1}$
and $\eta\eqref{chif2}\in S^{m_1+[m_2]-l+1,0}$ with
$\eta\partial_x\eqref{chif2}\in S^{m_1+[m_2]-l,0}$, which gives the thesis. \qed

\noindent{\it Proof of Lemma \ref{lemchi.3}}. The proof is based on
the fact that the flow of $h_1$ is essentially a rescaling of the flow
of $h_0$. Precisely, $\Phi^t_{h_1}$ leaves invariant the level
surfaces of $h_0$ and on a level surface $h_0=E$ one has
\begin{equation}
\label{fluh1}
\Phi^t_{h_1}\equiv \Phi^{(1+\epsilon f'(E))t}_{h_0}\ .
\end{equation}
So, we apply the formulae for the average and for $\chi$ getting the
result. We give the explicit proof of the fact that the solution
$\chi$ is a symbol. From \eqref{coh.2}
with $\Phi^t_{h_1}$ in place of  $\Phi^t_{h_0}$ we have
\begin{align*}
\chi&=\frac{1}{T_{h_1}}\int_0^{T_{h_1}} t\check p\circ \Phi_{h_1}^tdt 
= \frac{1}{(1+\epsilon f')^2 T_{h_1}}\int_0^{T_{h_1}} t(1+\epsilon
f')\check p\circ \Phi_{h_1}^t (1+\epsilon f')dt 
\\
&=\frac{1}{1+\epsilon f'} \frac{1}{ T_{h_0}}\int_0^{T_{h_0}} \tau
\check p\circ \Phi_{h_0}^\tau d\tau \ ,
\end{align*}
Now this is just $(1+\epsilon
f')^{-1}$ times the solution of the homological equation with the
original unperturbed Hamiltonian $h_0$. Since,  by the assumption  $(1+\epsilon
f')^{-1}$ is a symbol, which is a lower order
correction of the identity, the thesis follows. \qed

\subsection{Solution \eqref{12.1}}\label{cho3}

The homological equation \eqref{cho3} will be relevant only when
$l=1$, where we assume that $V(x)=x^2$ is a Harmonic potential. 

\begin{lemma}
\label{cohomo3} (Lemma 6.4 of \cite{Bam97})
The solution of the homological equation \eqref{cho3} is given by
$$ \chi(x,\xi,\phi):=\sum_{k\in\Z^n}\chi_k(x,\xi)e^{\ir k\cdot \phi}\ ,
$$
where
\begin{align}
\label{ch.3}
\chi_0=\frac{1}{T(E)}\int_0^{T(E)}t(\overline p-\overline{\langle
  p\rangle})\circ \Phi^t_{h_0}dt
\\
\label{ch.4}
\chi_k(x,\xi)=\frac{1}{e^{\ir\omega\cdot k
    T(E)}-1}\int_0^{T(E)}e^{\ir \omega\cdot k
  t}p_k(\Phi^t_{h_0}(x,\xi))dt \ ,
\end{align}
and $p_k$ is defined by 
$$
p_k(x,\xi):=\frac{1}{(2\pi)^{n}}\int_{\T^n}p(x,\xi,\phi)e^{-\ir k\phi}
d\phi\ .
$$
\end{lemma}

\begin{lemma}
\label{symb3}
Let $p\in S^{m_1,m_2}$, fix $\alpha\in\R$ and consider
\begin{equation}
\label{sym3.1}
I(x,\xi):=\int_0^{2\pi}e^{\im \alpha t}p\left(\Phi^t(x,\xi)\right)
dt\ .
\end{equation}
One has $I\in S^{m_1+[m_2],0}$ with $\partial_x p\in S^{m_1+[m_2]-1,0}$. 
\end{lemma}
\proof First we write the integral using the action angle variables
$(A,\theta)$ for the Harmonic oscillator. Thus we make the change of
variables
$$
x=\sqrt{A}\sin\theta\ ,\quad \xi=\sqrt{A}\cos\theta\ ;
$$
In these variables the flow is simply $\theta\to\theta+t $, so we have
\begin{align*}
I(A,\theta)=\int_0^{2\pi}e^{\im\alpha t}p_a(A,\theta+t)dt= e^{-\im
  \alpha\theta} \int_0^{2\pi}e^{\im\alpha t}p_a(A,t)dt
\\
= e^{-\im
  \alpha\theta} \int_0^{2\pi}e^{\im\alpha t}p(\sqrt{\xi^2+x^2}\cos t,
-\sqrt{\xi^2+x^2}\sin t) dt \ ,
\end{align*}

where $p_a(A,\theta)=p(\sqrt A\sin\theta,\sqrt A\cos \theta)$.

Now using a technique similar to that
used in the proof of Lemmas \ref{stint.6} and \ref{stixi}, one can see
that the integral is of class
$S^{m_1+[m_2]}$. 

In order to conclude the proof we have to check the prefactor. The
prefactor can be written as 
$$
\left(\frac{\xi-\im x}{A^{1/2}}\right)^\alpha\ ,
$$ which is easily seen to be a symbol which is bounded and has the
property that its $x$ derivative is bounded by $A^{-1/2}$, from which
the thesis immediately follows.\qed

\noindent{\it Proof of Lemma \ref{lemchi1}.} The result follows using
the previous Lemmas once
one has a lower bound of the small denominators. This is easily
obtained by remarking that, in
$ \Omega_{1\gamma}$ one has
\begin{align*}
\left|e^{\ir \omega\cdot
  kT}-1\right|=\left|2\sin\left(\frac{\omega\cdot kT}{2}\right)\right| 
\geq2 \left|\frac{\omega\cdot k T}{2}-k_0\pi\right| 
\\
=\left|\omega\cdot
k-k_0\right|\geq\frac{\gamma}{1+|k|^{\tau}}\ .
\end{align*}
\qed

\appendix

\section{Some technical lemmas}\label{lemms}

\begin{lemma}
\label{epM}
Let $f$ be a function of class $C^k$, and consider $f(1/E^{1/l})$. For
$E\to\infty$ one has:
\begin{equation}
\label{epM.1}
\frac{\partial^k}{\partial E^k}\left[f\left(\frac{1}{E^{1/l}}\right)\right] \asymp
\frac{1}{E^{k+\frac1l} }
\sum_{j=1}^{k}\frac{1}{E^{\frac{j-1}{l}}}f^{(j)}\left(\frac{1}{E^{\frac{1}{l}}}
\right)\ .
\end{equation}  
By $a\asymp b$ we mean $|a|\sleq |b|$ and $|b|\sleq |a|$, at least for
sufficiently large values of $E$.
\end{lemma}
\proof We use the Faa di Bruno formula which gives 
$$
\frac{\partial^k f}{\partial E^k}\asymp
\sum_{j=1}^{k}f^{(j)}(\mu)\sum_{h_1+...+h_j=k}
\frac{\partial^{h_1}\mu}{\partial E^{h_1}}
.... \frac{\partial^{h_j}\mu}{\partial E^{h_j}}\ ,
$$
where we denoted $\mu=E^{-1/l}$. The indexes $h_i$ always fulfill $h_i\geq
1$. On the other hand one has 
$$
\frac{\partial^{h}\mu}{\partial E^{h}} \asymp \frac{1}{E^{h+1/l}}\ ;
$$
substituting in the previous formula one gets the result. \qed 

\begin{lemma}
\label{stint}
Let $W(y,x)$ be a $C^\infty$ function fulfilling
\begin{equation}
\label{Wx}
\left|\partial^k_x W(y,x)\right|\sleq \langle x\rangle^{m-k}\ ,
\end{equation}
denote 
\begin{equation}
\label{stint.5}
I(M):=\int_{-1}^1\frac{W(y,My)}{\sq}dy
\end{equation}
then one has
\begin{equation}
\label{stint.1}
\left|\frac{\partial^k I}{\partial
  M^k} (M)  \right|\sleq \langle
M\rangle^{[m]-k}\ .
\end{equation}
\end{lemma}
\proof The difficulty in estimating the integral is that when $y=0$
the quantity $My$ does not diverge. One has 
\begin{equation}
\label{stint.2}
\frac{\partial^k}{\partial
  M^k}\int_{-1}^1\frac{W(y,My)}{\sq}dy=\int_{-1}^1 \frac{\partial_x^kW(y,My)y^k}
     {\sq}dy 
\end{equation}
We fix a small $a$ and split the interval of integration:
$[-1,1]=[-1,-a]\cup (-a,a)\cup [a,1]$. The integral over the first and
the last intervals are estimated in the same way. Consider the one
over $[a,1]$. One has
$$
\left|\int_a^1 \frac{\partial_x^kW(y,My)y^k}
     {\sq}dy \right|\leq \left|\int_a^1 \frac{\langle My\rangle^{m-k} y^k}
     {\sq}dy \right| \sleq \langle Ma\rangle^{m-k}\ .
$$
Over the interval $(-a,a)$ one has  $\sq>1/2$ provided $a$ is small
enough. Thus one has
\begin{align*}
 \left|\int_{-a}^a \frac{\partial_x^kW(y,My)y^k} {\sq}dy
\right|\sleq \int_{-a}^a {\langle My\rangle^{m-k} |y|^k} dy
=2\int_{0}^{Ma}\langle
q\rangle^{m-k}\left(\frac{q}{M}\right)^k\frac{dq}{M}
\\
=
\frac{2}{M^{k+1}} \int_{0}^{Ma}\langle
q\rangle^{m-k}{q}^k{dq}\sleq M^{[m]-k}\ ,
\end{align*}
which immediately gives the thesis.\qed

\begin{lemma}
\label{stint.6}
Under the same assumption of Lemma \ref{stint}, one has $I(E\bar q)\in
S^{[m]}$.
\end{lemma}
\proof First remark that, denoting $M=E^{\frac{1}{2l}}\bar q$, by
Lemma \ref{epM}, one has
$$
\partial_E^kM\asymp
\sum_{j=0}^{k}\partial^{k-j}_EE^{\frac{1}{2l}}\partial_E^{j}\bar q
\asymp \frac{E^{\frac{1}{2l}}}{E^{k}}\bar q+
\sum_{j=1}^{k}\frac{E^{\frac{1}{2l}}}{E^{k-j}}\frac{1}{E^{j+\frac{1}{l}}}\sum_{i=1}^{
j}\frac{\partial^i \bar q}{\partial
  \mu^i}\frac{1}{E^{\frac{i-1}{l}}}\asymp \frac{E^{1/2l}}{E^k}\ .  
$$
Now, from the Faa di Bruno formula one has
\begin{align*}
\partial^k_E
I(M)\asymp
\sum_{j=1}^{k}I^{(j)}(M)\sum_{h_1+...+h_j=k}\partial^{h_1}_MM...\partial^{h_j}_MM
\asymp \sum_{j=1}^{k}\langle M\rangle^{[m]-j}\sum_{h_1+...+h_j=k}
\frac{M}{E^{h_1}} ... \frac{M}{E^{h_j}}=\frac{M^{[m]}}{E^k}\ ,
\end{align*}
from which the thesis follows. \qed

By working as in the proof of the above lemmas one gets also the
following useful result.

\begin{lemma}
\label{stixi}
Let $g(y,\xi)$ be such that
$$
\left|\partial_\xi^kg(x,\xi)\right|\sleq \lambda^{m-kl}\ ,
$$
consider 
$$
I(E):=\int_{-1}^1\frac{g(y,\sqrt{E-V(q(y))})}{\sq }dy\ ,
$$
then one has $I\in S^m$.
\end{lemma}


\end{document}